\documentclass[12pt]{article}
\usepackage{epsf}
\title{ {\bf
Electric dipole moments of charged leptons in the split fermion
scenario in the two  Higgs doublet model.}}
\author{\vspace{1cm}\\
        {\bf E. O. Iltan}
        \thanks{E-mail address:
        eiltan@heraklit.physics.metu.edu.tr}
 \\
        Physics Department, Middle East Technical University \\
        Ankara, Turkey\\}

\date{}

\begin{document}
\setlength{\baselineskip}{24pt}
\maketitle
\setlength{\baselineskip}{7mm}
\begin{abstract}
We predict the charged lepton electric dipole moments in the split
fermion scenario in the framework of the two Higgs doublet model.
We observe that the numerical value of the muon (tau)  electric
dipole moment is at the order of the magnitude of $10^{-22}\,
(e-cm)$ ($10^{-20}\, (e-cm)$) and there is an enhancement in the
case of two extra dimensions, especially for the tau lepton
electric dipole moment.
\end{abstract}
\thispagestyle{empty}
\newpage
\setcounter{page}{1}
\section{Introduction}
The existence of the electric dipole moments (EDMs) of fermions
depends on the CP violating interactions. The complex Cabibo
Kobayashi Maskawa (CKM) matrix elements cause the CP violation in
the standard model (SM), however, the estimated fermion EDMs are
negligibly small, if the above complex phases are considered. This
stimulates one to investigate these physical quantities in the
framework of the new models beyond the SM, such as multi Higgs
doublet models (MHDM), supersymmetric model (SUSY),
\cite{Schmidt},..., etc..

In the literature there exist the experimental results on the
fermion EDMs, $d_e =(1.8\pm 1.2\pm 1.0)\times 10^{-27} e\, cm$
\cite{Commins}, $d_{\mu} =(3.7\pm 3.4)\times 10^{-19} e\, cm$
\cite{Bailey} and $d_{\tau} =(3.1)\times 10^{-16} e\, cm$
\cite{Groom} respectively and on the neutron EDM h $d_N <
1.1\times 10^{-25} e\, cm$ \cite{Smith1}.

There is an extensive theoretical work done on the EDMs of
fermions. The quark EDMs have been estimated in several models
\cite{sahab1} and the EDMs of nuclei, deutron, neutron and some
atoms have been studied extensively \cite{Vladimir}. The lepton
electric dipole moments have been predicted in various studies
\cite {Bhaskar}-\cite {IltanExtrEDM}. In \cite {Bhaskar} the
lepton electric dipole moments  has been analyzed in the framework
of the seesaw model. \cite {Iltmuegam} was devoted to the EDMs of
the leptons in the model III version of the 2HDM and $d_e$ has
been predicted at the order of the magnitude of $10^{-32}\, e-cm$.
The work \cite{IltanNonCom} was related to the lepton EDM moments
in the framework of the SM with the inclusion of non-commutative
geometry. Furthermore,  the effects of non-universal extra
dimensions on the electric dipole moments of fermions in the two
Higgs doublet model have been estimated in \cite{IltanExtrEDM}.

This work is devoted to the prediction of the lepton EDMs in the
two Higgs doublet model in which the flavor changing (FC) neutral
current vertices in the tree level is permitted and the CP
violating interactions are carried by complex Yukawa couplings.
Furthermore, we respect the split fermion scenario where the
hierarchy of fermion masses are coming from the overlap of the
fermion Gaussian profiles in the extra dimensions. The split
fermion scenario has been studied in several works in the
literature \cite{Hamed}-\cite{Delgado}. In \cite{Hamed} the
alternative view of the fermion mass hierarchy has been introduced
by assuming that the fermions were located at different points in
the extra dimensions and this geometric interpretation resulted in
the exponentially small overlaps of their wavefunctions. The
separation of fermions in the extra dimension forbids the local
couplings between quarks and leptons and this can ensure a
solution also to the proton stability. \cite{Mirabelli} was
devoted to the locations of left and right handed components of
fermions in the extra dimensions and their roles on the mechanism
of Yukawa hierarchies. The constraint on the split fermions in the
extra dimensions has been obtained by considering leptonic W
decays and the lepton violating processes in \cite{Chang}. The
discussion of CP violation in the quark sector in the split
fermion model was done in \cite{Branco}. \cite{Chang2} was related
to the new configuration of split fermion positions in a single
extra dimension and the physics of kaon, neutron and B/D mesons to
find stringent bounds on the size of the compactification scale
1/R. The contributions due to the split fermion scenario on the
rare processes has been examined in \cite{Hewett} and the shapes
and overlaps of the fermion wave functions in the split fermion
model have been studied in \cite{Perez}.

In the present work, we consider the EDMs of charged leptons by
assuming that they have Gaussian profiles in the extra dimensions.
First, we study the EDMs of charged leptons in a single extra
dimension, using the estimated location of them. Then, we assume
that the number of extra dimensions is two and the charged leptons
are restricted to the fifth extra dimension, with non-zero
Gaussian profiles. As a final analysis, we predict the EDMs of
charged leptons by taking non-zero Gaussian profiles in both extra
dimensions.

In the numerical calculations, we observe that the electron  EDM
$d_e$ is at the order of $10^{-32}$ e-cm and it is too small to
study the additional effects due to the extra dimensions. The
numerical value of $d_{\mu}$ ($d_{\tau}$) is at most at the order
of the magnitude of $10^{-22}\, (10^{-20})$ e-cm in the case that
the leptons have non-zero Gaussian profiles in the first extra
dimension, for two extra dimensions, for the large values of the
compactification scale $1/R$ and for the intermediate values of
Yukawa couplings.

The paper is organized as follows: In Section 2, we present EDMs
of charged leptons in the split fermion scenario, in the two Higgs
doublet model. Section 3 is devoted to discussion and our
conclusions.
\section{Electric dipole moments of charged leptons in the split fermion
scenario in the two  Higgs doublet model }
The fermion EDM is carried by the CP violating
fermion-fermion-photon effective interaction and, for quarks (for
charged leptons), the complex CKM matrix  (possible lepton mixing
matrix) elements is the possible source of this violation, in the
framework of the SM. The estimated tiny numerical values of EDMs
of fermions in the SM stimulates one to go beyond and the model
III version of the 2HDM is one of the candidate to get relatively
large EDM values, since the FC neutral currents (FCNC) are
permitted at tree level and the new Yukawa couplings can be
complex in general. Furthermore, the addition of the spatial extra
dimensions brings additional contributions sensitive to the
compactification scale $1/R$ where $R$ is the radius of the
compactification. Here we take the effects of extra dimensions
into account and we follow the idea that the hierarchy of lepton
masses are coming from the lepton Gaussian profiles in the extra
dimensions.

The Yukawa Lagrangian responsible for the lepton EDM in a single
extra dimension, respecting the split fermion scenario, reads:
\begin{eqnarray}
{\cal{L}}_{Y}=
\xi^{E}_{5 \,ij} \bar{\hat{l}}_{i L} \phi_{2} \hat{E}_{j R} + h.c.
\,\,\, , \label{lagrangian}
\end{eqnarray}
where $L$ and $R$ denote chiral projections $L(R)=1/2(1\mp
\gamma_5)$, $\phi_{2}$ is the new scalar doublet. Here $\hat{l}_{i
L}$ ($\hat{E}_{j R}$), with family indices $i,j$, are the zero
mode \footnote{In our calculations, we take only zero mode lepton
fields. See Discussion section for further explanation.} lepton
doublets (singlets) with Gaussian profiles in the extra dimension
$y$ and they read
\begin{eqnarray}
\hat{l}_{i L}&=& N\,e^{-(y-y_{i L})^2/2 \sigma^2}\,l_{i L} ,
\nonumber
\\ \hat{E}_{j R}&=&N\, e^{-(y-y_{j R})^2/2 \sigma^2}\, E_{j R}\, ,
\label{gaussianprof}
\end{eqnarray}
with the normalization factor $N=\frac{1}{\pi^{1/4}\,
\sigma^{1/2}}$. $l_{i L}$ ($E_{j R}$) are the lepton doublets
(singlets) in four dimension. The parameter $\sigma$ is the
Gaussian width of the leptons with the property $\sigma << R$,
$y_{i (L,R)}$ are the fixed position of $i^{th}$ left (right)
handed lepton in the fifth dimension. The positions of left handed
and right handed leptons are obtained by taking the observed
masses into account \cite{Mirabelli}. The idea is that the lepton
mass hierarchy is due to the the relative positions of the
Gaussian peaks of the wave functions located in the extra
dimension \cite{Hamed, Mirabelli}. By assuming that the lepton
mass matrix is diagonal, one possible set of locations for the
lepton fields read (see \cite{Mirabelli} for details)
\begin{eqnarray}
P_{l_i}=\sqrt{2}\,\sigma\, \left(\begin{array}{c c c}
11.075\\1.0\\0.0
\end{array}\right)\,,\,\,\,\, P_{e_i}=\sqrt{2}\,\sigma\, \left(\begin{array}
{c c c} 5.9475\\4.9475\\-3.1498
\end{array}\right)
 \,\, . \label{location}
\end{eqnarray}

Here we choose the Higgs doublets $\phi_{1}$ and $\phi_{2}$ as
\begin{eqnarray}
\phi_{1}=\frac{1}{\sqrt{2}}\left[\left(\begin{array}{c c}
0\\v+H^{0}\end{array}\right)\; + \left(\begin{array}{c c} \sqrt{2}
\chi^{+}\\ i \chi^{0}\end{array}\right) \right]\, ;
\phi_{2}=\frac{1}{\sqrt{2}}\left(\begin{array}{c c} \sqrt{2}
H^{+}\\ H_1+i H_2 \end{array}\right) \,\, . \label{choice}
\end{eqnarray}
with the vacuum expectation values,
\begin{eqnarray}
<\phi_{1}>=\frac{1}{\sqrt{2}}\left(\begin{array}{c c}
0\\v\end{array}\right) \,  \, ; <\phi_{2}>=0 \,\, .
\label{choice2}
\end{eqnarray}
and collect SM (new) particles in the first (second) doublet.
Notice that $H_1$ and $H_2$ are the mass eigenstates $h^0$ and
$A^0$ respectively since no mixing occurs between two CP-even
neutral bosons $H^0$ and $h^0$ at tree level, in our case.

\begin{figure}[htb]
\vskip 0.5truein \centering \epsfxsize=2.8in
\leavevmode\epsffile{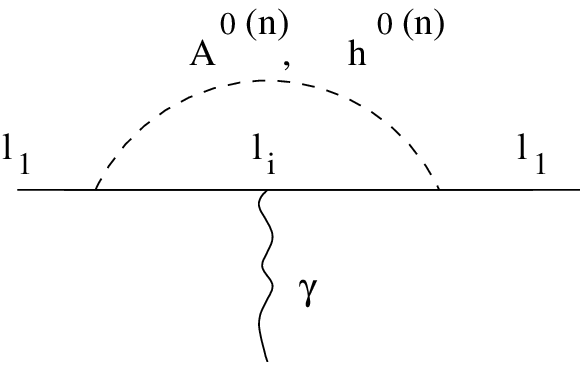} \vskip 0.5truein \caption[]{One loop
diagrams contribute to EDM of charged leptons due to neutral Higgs
bosons $h^0$, $A^0$ in the 2HDM, including KK modes in a single
extra dimension. Wavy lines represent the electromagnetic field
and dashed lines the Higgs field where $l_{1 \,(i)}=e, \mu, \tau$}
\label{fig1}
\end{figure}

The new Higgs field $\phi_{2}$, playing the main role in the
existence of the charged lepton EDM, is accessible to extra
dimension and after the compactification on the orbifold
$S^1/Z_2$, it is expanded as
\begin{eqnarray}
\phi_{2}(x,y ) & = & {1 \over {\sqrt{2 \pi R}}} \left\{
\phi_{2}^{(0)}(x) + \sqrt{2} \sum_{n=1}^{\infty} \phi_{2}^{(n)}(x)
\cos(ny/R)\right\} \,, \label{SecHiggsField}
\end{eqnarray}
where $\phi_{2}^{(0)}(x)$ is the 4-dimensional Higgs doublet which
contains the charged Higgs boson $H^+$, the neutral CP even (odd)
Higgs bosons $h^0$ ($A^0$) and $\phi_{2}^{(n)}(x)$ are the KK
modes of $\phi_{2}$. The non-zero KK mode of Higgs doublet
$\phi_{2}$ includes a charged Higgs of mass
$\sqrt{m_{H^\pm}^2+m_n^2}$, a neutral CP even Higgs of mass
$\sqrt{m_{h^0}^2+m_n^2}$, a neutral CP odd Higgs of mass
$\sqrt{m_{A^0}^2+m_n^2}$ where $m_n=n/R$ is the mass of $n$'th
level KK particle. Notice that the gauge field KK modes do not
bring new contributions to the EDMs of charged leptons since they
do not exist in the one loop diagrams \\ (see Fig. \ref{fig1}).
Now, we present the vertices existing in the diagrams, after the
integration over the fifth dimension. The integration of the
combination $\bar{\hat{f}}_{iL\,(R)}\,S^{(n)}(x)\,\cos(ny/R)\,
\hat{f}_{j R\,(L)}$, ($S=h^0, A^0$), appearing in the part of the
Lagrangian (eq. (\ref{lagrangian})), over the fifth dimension
reads
\begin{eqnarray}
\int_{-\pi R}^{\pi R}\, dy\,\,
\bar{\hat{f}}_{iL\,(R)}\,S^{(n)}(x)\,\cos(ny/R)\,
\hat{f}_{jR\,(L)}=V^n_{LR\,(RL)\,ij} \, \bar{f}_{i L\, (R)}
\,S^{(n)} (x)\,\,f_{j R\,(L)}\,\, , \label{intVij1}
\end{eqnarray}
where the factor $V^n_{LR\,(RL)\,ij}$ is
\begin{eqnarray}
V^n_{LR\,(RL)\,ij}=e^{-n^2\,\sigma^2/4\,R^2}\,e^{-(y_{i L\,
(R)}-y_{j R\, (L)})^2/4 \sigma^2}\, \cos\, [\frac{n\,(y_{i L\,
(R)}+y_{j R\, (L)})}{2\,R}] \,\, . \label{Vij1}
\end{eqnarray}
Here the fields $f_{iL}$, $f_{j R}$ are the four dimensional
lepton fields. Therefore we can define the Yukawa couplings in
four dimensions as
\begin{eqnarray}
\xi^{E}_{ij}\,\Big((\xi^{E \dagger}_{ij})^\dagger\Big)=
V^0_{LR\,(RL)\,ij} \, \xi^{E}_{5\, ij}\,\Big((\xi^{E}_{5\,
ij})^\dagger\Big)/\sqrt{2 \pi R} \,\, . \label{coupl4}
\end{eqnarray}
where $\xi^{E}_{5\, ij}$ are  Yukawa couplings in five dimensions
(see eq. (\ref{lagrangian}))
\footnote{In the following we use the dimensionful coupling
$\bar{\xi}^{E}_{N}$ with the definition
$\xi^{E}_{N,ij}=\sqrt{\frac{4\, G_F}{\sqrt{2}}}\,
\bar{\xi}^{E}_{N,ij}$ where N denotes the word "neutral".}.

In the case of two extra dimensions, after the compactification on
the orbifold $(S^1\times S^1)/Z_2$, the new Higgs field $\phi_{2}$
can be expanded as
\begin{eqnarray}
\phi_{2}(x,y,z ) & = & {1 \over {2 \pi R}} \left\{
\phi_{2}^{(0,0)}(x) + 2 \sum_{n,s}^{\infty} \phi_{2}^{(n,s)}(x)
\cos(ny/R+sz/R)\right\} \,, \label{SecHiggsField2}
\end{eqnarray}
where $\phi_{2}^{(n,s)}(x)$ are the KK modes of $\phi_{2}$. Notice
that the mass of KK modes of the charged (neutral CP even, neutral
CP odd) Higgs  is $\sqrt{m_{H^\pm}^2+m_n^2+m_s^2}$,
($\sqrt{m_{h^0}^2+m_n^2+m_s^2}$, \\ $\sqrt{m_{A^0}^2+m_n^2+m_s^2}$
) where $m_n=n/R \, (m_s=s/R)$ is the mass of $n (s)$'th level KK
particle. In the case that the leptons are restricted only the
fifth dimension, the vertex factor $V^{n}_{LR\,(RL)\,ij}$ is the
same as the one in eq. (\ref{Vij1}). On the other hand, if we
assume that the leptons are accessible two both dimensions with
Gaussian profiles as
\begin{eqnarray}
\hat{l}_{i L}&=& N\,e^{-\Big((y-y_{i L})^2+(z-z_{i L})^2\Big)/2
\sigma^2}\,l_{i L} , \nonumber
\\ \hat{E}_{j R}&=&N\, e^{-\Big((y-y_{j R})^2+(z-z_{j R})^2\Big)/2
\sigma^2}\, E_{j R}\, , \label{gaussianprof2}
\end{eqnarray}
with the normalization factor $N=\frac{1}{\pi^{1/2}\, \sigma}$,
the integration of the part of the Lagrangian \\
$\bar{\hat{f}}_{iL\,(R)}\,S^{(n,s)}(x)\,\cos(ny/R+sz/R)\,
\hat{f}_{j R\,(L)}$ over the fifth and sixth extra dimensions
results in the vertex factor

\begin{eqnarray}
V^{(n,s)}_{LR\,(RL)\,ij}&=&e^{-(n^2+s^2)\,\sigma^2/4\,R^2}\,e^{-\Big(
(y_{i L\, (R)}-y_{j R\, (L)})^2+(z_{i L\, (R)}-z_{j R\, (L)})^2
\Big)/4 \sigma^2}\nonumber \\&\times& \cos\, [\frac{n\,(y_{i L\,
(R)}+y_{j R\, (L)})+s\,(z_{i L\, (R)}+z_{j R\, (L)})}{2\,R}] \,.
\label{Vij2}
\end{eqnarray}
Similar to a single extra dimension case, we define the Yukawa
couplings in four dimension as
\begin{eqnarray}
\xi^{E}_{ij}\,\Big((\xi^{E}_{ij})^\dagger\Big)=
V^{(0,0)}_{LR\,(RL)\,ij} \, \xi^{E}_{6 \,ij}\,\Big((\xi^{E}_{6
\,ij})^\dagger\Big)/2 \pi R\, , \label{coupl44}
\end{eqnarray}
where $V^{(0,0)}_{LR\,(RL)\,ij}=e^{-\Big( (y_{i L\, (R)}-y_{j R\,
(L)})^2+(z_{i L\, (R)}-z_{j R\, (L)})^2 \Big)/4 \sigma^2}$. Here,
we present a possible positions of left handed and right handed
leptons in the two extra dimensions, by respecting the observed
masses \footnote{The calculation is similar to the one presented
in \cite{Mirabelli} which is done for a single extra dimension.}.
Similar to the previous discussion, we assume that the lepton mass
matrix is diagonal and one of the possible set of locations for
the Gaussian peaks of the lepton fields in the two extra
dimensions read
\begin{eqnarray}
P_{l_i}=\sqrt{2}\,\sigma\, \left(\begin{array}{c c c}
(8.417,8.417)\\(1.0,1.0)\\(0.0,0.0)
\end{array}\right)\,,\,\,\,\,
P_{e_i}=\sqrt{2}\,\sigma\, \left(\begin{array} {c c c}
(4.7913,4.7913)\\(3.7913,3.7913)\\(-2.2272,-2.2272)
\end{array}\right)
 \,\, . \label{location2}
\end{eqnarray}
where the numbers in the parenthesis denote the y and z
coordinates of the location of the Gaussian peak in the extra
dimensions. Here we choose the same numbers for the y and z
locations of the Gaussian peaks.

Now, we would like to present the EDMs of charged leptons with the
addition of a single extra dimension where the localized leptons
have Gaussian profiles. The effective EDM interaction for a
charged lepton $f$ is given by
\begin{eqnarray}
{\cal L}_{EDM}=i d_f \,\bar{f}\,\gamma_5 \,\sigma^{\mu\nu}\,f\,
F_{\mu\nu} \,\, , \label{EDM1}
\end{eqnarray}
where $F_{\mu\nu}$ is the electromagnetic field tensor, '$d_{f}$'
is EDM of the charged lepton and it is a real number by
hermiticity. With the assumption that there is no CKM type lepton
mixing matrix and ignoring possible CP violating LFV interactions
due to the lepton-lepton KK mode-Higgs KK mode vertices, only the
new neutral Higgs part gives a contribution to their EDMs and
$f$-lepton EDM '$d_f$' $(f=e,\,\mu,\,\tau)$  can be calculated as
a sum of contributions coming from neutral Higgs bosons $h_0$ and
$A_0$,
\begin{eqnarray}
d_f&=& -\frac{i G_F}{\sqrt{2}} \frac{e}{32\pi^2}\,
\frac{Q_{\tau}}{m_{\tau}}\, ((\bar{\xi}^{D\,*}_{N,l\tau})^2-
(\bar{\xi}^{D}_{N,\tau l})^2)\, \Bigg ( (F_1 (y_{h_0})-F_1
(y_{A_0}))\nonumber \\ &+& 2\,
\sum_{n=1}^{\infty}\,e^{-n^2\,\sigma^2/2\,R^2}\,c_n (f,\tau)\,c'_n
(f,\tau)\, (F_1 (y^n_{h_0})-F_1 (y^n_{A_0})) \Bigg)\, ,
\label{emuEDM}
\end{eqnarray}
for $f=e,\mu$ and
\begin{eqnarray}
d_{\tau}&=& -\frac{i G_F}{\sqrt{2}} \frac{e}{32\pi^2}\, \Big{\{}
\frac{Q_{\tau}}{m_{\tau}}\, ((\bar{\xi}^{D\,*}_{N,\tau\tau})^2-
(\bar{\xi}^{D}_{N,\tau \tau})^2)\, \Bigg( (F_2
(r_{h_0})-F_2(r_{A_0}))\nonumber \\ &+& 2\,
\sum_{n=1}^{\infty}\,e^{-n^2\,\sigma^2/2\,R^2} \,c^2_n
(\tau,\tau)\,(F_2 (r^n_{h_0})-F_2(r^n_{A_0}))\Bigg) \nonumber
\\&-&  Q_{\mu}\, \frac{m_{\mu}}{m^2_{\tau}}\,
((\bar{\xi}^{D\,*}_{N,\mu\tau})^2-(\bar{\xi}^{D}_{N,\tau
\mu})^2)\, \Bigg((r_{h_0}\,ln\, (z_{h_0})-r_{A_0}\,ln\,
(z_{A_0}))\nonumber \\ &+& 2\,
\sum_{n=1}^{\infty}\,e^{-n^2\,\sigma^2/2\,R^2}\,c_n
(\mu,\tau)\,c'_n (\mu,\tau)\,  (r^n_{h_0}\,ln\,
(z^n_{h_0})-r^n_{A_0}\,ln\, (z^n_{A_0}))\Bigg) \Big{\}} \,\, ,
\label{tauEDM}
\end{eqnarray}
where
\begin{eqnarray}
c_n \,(f,\tau)&=&\cos[\frac{n\,(y_{f R}+y_{\tau L})}{2\, R}]\, \,, \nonumber \\
c'_n \,(f,\tau)&=&\cos[\frac{n\,(y_{f L}+y_{\tau R})}{2\, R}]\, ,
\label{coeff}
\end{eqnarray}
for $f=e,\mu, \tau$ and the functions $F_1 (w)$, $F_2 (w)$ read
\begin{eqnarray}
F_1 (w)&=&\frac{w\,(3-4\,w+w^2+2\,ln\,w)}{(-1+w)^3}\nonumber \,\, , \\
F_2 (w)&=& w\, ln\,w + \frac{2\,(-2+w)\, w\,ln\,
\frac{1}{2}(\sqrt{w}-\sqrt{w-4})}{\sqrt{w\,(w-4)}} \, ,
\label{functions1}
\end{eqnarray}
with $y^n_{S}=\frac{m^2_{\tau}}{m^2_{S}+n^2/R^2}$,
$r^n_{S}=\frac{1}{y^n_{S}}$ and
$z^n_{S}=\frac{m^2_{\mu}}{m^2_{S}+n^2/R^2}$, $y_{S}=y^0_{S}$,
$r_{S}=r^0_{S}$ and $z_{S}=z^0_{S}$, $Q_{\tau}$, $Q_{\mu}$ are
charges of $\tau$ and $\mu$ leptons respectively. In eq.
(\ref{emuEDM}) we take into account only internal $\tau$-lepton
contribution respecting our assumption that the Yukawa couplings
$\bar{\xi}^{E}_{N, ij},\, i,j=e,\mu$, are small compared to
$\bar{\xi}^{E}_{N,\tau\, i}\, i=e,\mu,\tau$ due to the possible
proportionality of the Yukawa couplings to the masses of leptons
in the vertices. In eq. (\ref{tauEDM}) we present also the
internal $\mu$-lepton contribution, which can be neglected
numerically. Notice that, we make our calculations in arbitrary
photon four momentum square $q^2$ and take $q^2=0$ at the end.

Now, we present the EDMs of charged leptons with the addition of
two extra dimensions for the case that the leptons are accessible
to both extra dimensions:
\begin{eqnarray}
d_f&=& -\frac{i G_F}{\sqrt{2}} \frac{e}{32\pi^2}\,
\frac{Q_{\tau}}{m_{\tau}}\, ((\bar{\xi}^{D\,*}_{N,l\tau})^2-
(\bar{\xi}^{D}_{N,\tau l})^2)\, \Bigg ( (F_1 (y_{h_0})-F_1
(y_{A_0}))\nonumber \\ &+& 4\,
\sum_{n,s}^{\infty}\,e^{-(n^2+s^2)\,\sigma^2/2\,R^2}\,c_{2\,(n,s)}
(f,\tau)\,c'_{2\,(n,s)} (f,\tau)\, (F_1 (y^{(n,s)}_{h_0})-F_1
(y^{(n,s)}_{A_0})) \Bigg)\, , \label{emuEDM2}
\end{eqnarray}
for $f=e,\mu$ and
\begin{eqnarray}
d_{\tau}&=& -\frac{i G_F}{\sqrt{2}} \frac{e}{32\pi^2}\, \Big{\{}
\frac{Q_{\tau}}{m_{\tau}}\, ((\bar{\xi}^{D\,*}_{N,\tau\tau})^2-
(\bar{\xi}^{D}_{N,\tau \tau})^2)\, \Bigg( (F_2
(r_{h_0})-F_2(r_{A_0}))\nonumber \\ &+& 4\,
\sum_{n,s}^{\infty}\,e^{-(n^2+s^2)\,\sigma^2/2\,R^2}
\,(c_{2\,(n,s)})^2 (\tau,\tau)\,(F_2
(r^(n,s)_{h_0})-F_2(r^(n,s)_{A_0}))\Bigg) \nonumber
\\&-&  Q_{\mu}\, \frac{m_{\mu}}{m^2_{\tau}}\,
((\bar{\xi}^{D\,*}_{N,\mu\tau})^2-(\bar{\xi}^{D}_{N,\tau
\mu})^2)\, \Bigg((r_{h_0}\,ln\, (z_{h_0})-r_{A_0}\,ln\,
(z_{A_0}))\nonumber \\ &+&\!\!\!\! 4\,
\sum_{n,s}^{\infty}\,e^{-(n^2+s^2)\,\sigma^2/2\,R^2}\,c_{2\,(n,s)}
(\mu,\tau)\,c'_{2\,(n,s)} (\mu,\tau)\,  (r^{(n,s)}_{h_0}\,ln\,
(z^{(n,s)}_{h_0})-r^{(n,s)}_{A_0}\,ln\, (z^{(n,s)}_{A_0}))\Bigg)\!
\Big{\}}, \label{tauEDM2}
\end{eqnarray}
where
\begin{eqnarray}
c_{2\,(n,s)}\, (f,\tau)&=&\cos[\frac{n\,(y_{f R}+y_{\tau L})+
s\,(z_{f R}+z_{\tau L})}{2\, R}]\, \,,
\nonumber \\
c'_{2\,(n,s)}(f,\tau)&=&\cos[\frac{n\,(y_{f L}+y_{\tau R}) +
s\,(z_{f L}+z_{\tau R})}{2\, R}]\, , \label{coeff2}
\end{eqnarray}
for $f=e,\mu, \tau$. In eqs. (\ref{emuEDM2}) and (\ref{tauEDM2}),
the parameters $y^{(n,s)}_{S}$, $r^{(n,s)}_S$ and $z^{(n,s)}_{S}$
are defined as,
$y^{(n,s)}_{S}=\frac{m^2_{\tau}}{m^2_{S}+n^2/R^2+s^2/R^2}$,
$r^{(n,s)}_{S}=\frac{1}{y^{(n,s)}_{S}}$ and
$z^{(n,s)}_{S}=\frac{m^2_{\mu}}{m^2_{S}+n^2/R^2+s^2/R^2}$. In eqs.
(\ref{emuEDM2}) and (\ref{tauEDM2}) the summation would be done
over $n,s=0,1,2 ...$ except $n=s=0$.

Finally, in our calculations, we choose the Yukawa couplings
complex and we used the parametrization
\begin{eqnarray}
\bar{\xi}^{E}_{N,\tau f}=|\bar{\xi}^{E}_{N,\tau f}|\, e^{i\,
\theta_{f}} \,. \label{xi2}
\end{eqnarray}
Therefore, the Yukawa factors in eqs. (\ref{emuEDM}),
(\ref{tauEDM}), (\ref{emuEDM2}) and (\ref{tauEDM2}) can be written
as
\begin{eqnarray}
((\bar{\xi}^{D\,*}_{N,f\tau})^2-(\bar{\xi}^{D}_{N,\tau
f})^2)=-2\,i \,sin\,2\theta_{f}\, |\bar{\xi}^{D}_{N,\tau f}|^2
\end{eqnarray}
where $f=e,\mu,\tau$. Here $\theta_{f}$ is the CP violating
parameter which is the source of the lepton EDM.

\section{Discussion}
This work is devoted to the analysis  of the effects extra
dimensions on the EDMs of fermions  in the case that the hierarchy
of lepton masses are due to the lepton Gaussian profiles in the
extra dimensions. The CP violating nature of the EDM interactions
needs the CP violating phases. Here, for the complex phases, we
consider the complex Yukawa couplings appearing in the FCNC at
tree level in the framework of the 2HDM. Notice that we do not
take the internal lepton KK modes into account (see for example
\cite{Mirabelli} for the calculation of the KK modes of leptons).
The lepton-lepton KK mode-Higgs zero mode and lepton-lepton KK
mode-Higgs KK mode vertices carry the possible LFV and CP
violating interactions. We ignored these contributions because of
the difficulty arising during the summations. We expect that, for
the large values of the compactification scale, their effects on
the physical parameters are suppressed.

The four dimensional leptonic complex couplings
$\bar{\xi}^E_{N,ij}, i,j=e, \mu, \tau$ are the free parameters of
2HDM. We consider the Yukawa couplings $\bar{\xi}^{E}_{N,ij},\,
i,j=e,\mu $, as smaller compared to $\bar{\xi}^{E}_{N,\tau\, i}\,
i=e,\mu,\tau$ and we assume that $\bar{\xi}^{E}_{N,ij}$ is
symmetric with respect to the indices $i$ and $j$. In the case
that no extra dimension exists, the upper limit of
$\bar{\xi}^{D}_{N,\tau \mu}$ is predicted as $30\, GeV$ (see
\cite{Iltananomuon} and references therein) by using the
experimental uncertainty, $10^{-9}$, in the measurement of the
muon anomalous magnetic moment \cite{BNL} and assuming that the
new physics effects can not exceed this uncertainty. Using this
upper limit and the experimental upper bound of BR of
$\mu\rightarrow e \gamma$ decay, BR $\leq 1.2\times 10^{-11}$, the
coupling $\bar{\xi}^{D}_{N,\tau e}$ can be restricted in the
range, $10^{-3}-10^{-2}\, GeV$ (see \cite{Iltmuegam}). In our
calculations we choose the numerical values of the couplings
$\bar{\xi}^{E}_{N,\tau \mu}$ ($\bar{\xi}^{E}_{N,\tau e}$) around
$30\, GeV$ ($10^{-2}\, GeV$). For the coupling
$\bar{\xi}^{E}_{N,\tau \tau}$, we use the numerical values which
are greater than $\bar{\xi}^{E}_{N,\tau \mu}$, since we have no
explicit restriction region.

Here, we respect the split fermion scenario where the hierarchy of
lepton masses are due to the lepton Gaussian profiles in the extra
dimensions. The SM scalar $H^0$ is a constant profile in the extra
dimensions and the mass term, which is modulated by the mutual
overlap of lepton wavefunctions, is obtained by integrating the
operator $H^0\,\bar{\hat{f}}\,\bar{f}$ over extra dimensions. This
idea is the main point to fix the position of left (right) handed
lepton in the extra dimensions (see \cite{Mirabelli} for details).
Since the leptons are located in the extra dimensions with
Gaussian profiles, the parameter $\rho=\sigma/R$, where $\sigma$
is the Gaussian width of the fermions, is the free parameter of
the model. The locations of lepton fields in the extra dimensions
are obtained in terms of Gaussian width $\sigma$.

In the present work we take split leptons in a single and two
extra dimensions and use a possible set of locations to calculate
the strength of the lepton-lepton-new Higgs scalars vertices,
which play the main role in the calculation of the charged lepton
EDM. First, we take a single extra dimension and use the estimated
location of the leptons (see eq. (\ref{location})) to calculate
the corresponding vertices (eq. (\ref{intVij1})). After that, we
assume that the number of extra dimensions is two and take that
the leptons are restricted to the fifth extra dimension, with
non-zero Gaussian profiles. In this case the abundance of new
scalar KK modes causes to increase the EDM of charged leptons,
especially the $\tau$ lepton EDM. However, the exponential
suppression factor (see eq. (\ref{Vij1})) appearing in the
summation of the KK modes causes that the sum does not have a
large contribution. Finally, we assume that the leptons have
non-zero Gaussian profiles also in the sixth dimension and using a
possible set of locations in the fifth and sixth extra dimensions
(see eq. (\ref{location2})), we calculated the EDM of charged
leptons. In this case the additional the exponential factor
appearing in the second summation further suppresses the lepton
EDM especially for the muon case.

Now, we start to estimate the charged lepton EDMs and to study the
parameter $\rho$ and the compactification scale $1/R$ dependencies
of these measurable quantities.

\begin{figure}[htb]
\vskip -3.0truein \centering \epsfxsize=6.8in
\leavevmode\epsffile{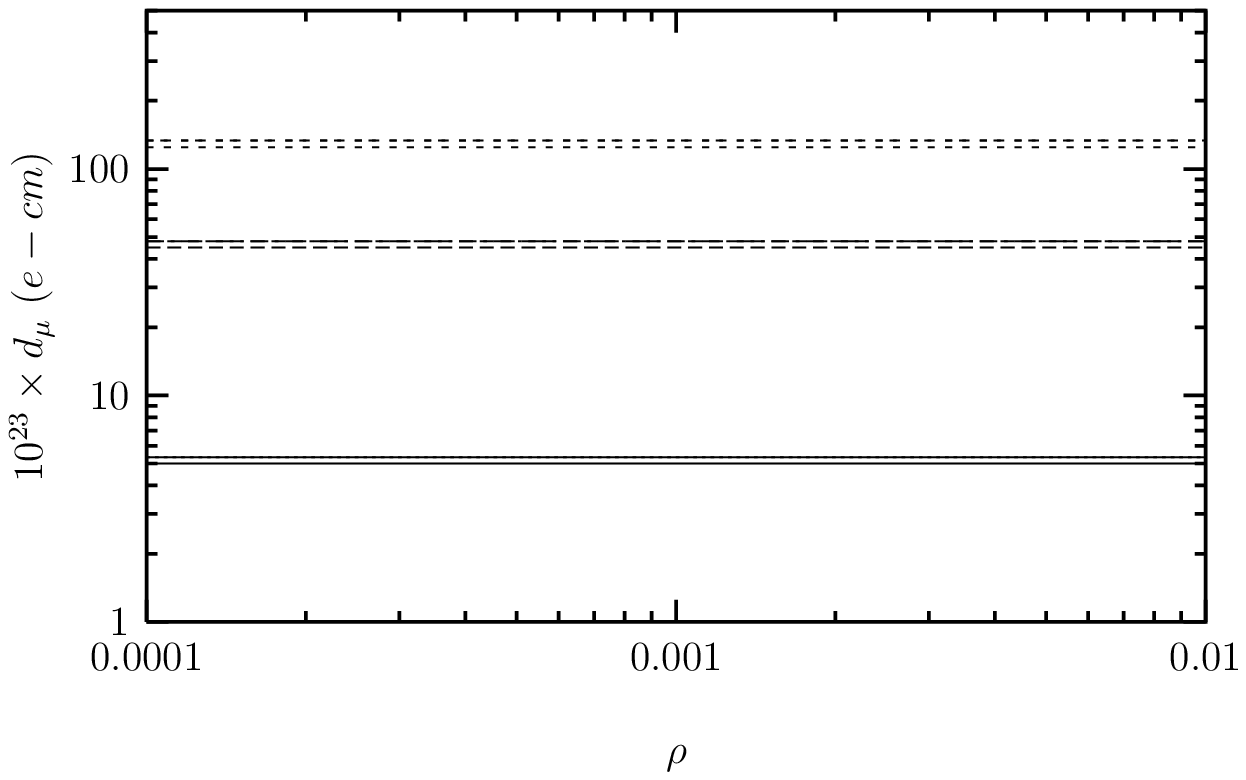} \vskip -3.0truein
\caption[]{$d_{\mu}$ with respect to the parameter $\rho$ for
$1/R=500\,GeV$, $m_{h^0}=100\, GeV$, $m_{A^0}=200\, GeV$ and the
intermediate value of $sin\,\theta_{\mu}=0.5$. The lower-upper
solid (dashed, small dashed) line represents the $d_{\mu}$ for a
single-two extra dimensions, for $\bar{\xi}^{E}_{N,\tau \mu}
=10\,(30,\, 50)\, GeV$.} \label{EDMmu500ro}
\end{figure}

In  Fig. \ref{EDMmu500ro}, we plot EDM $d_{\mu}$ with respect to
the parameter $\rho$ for $1/R=500\,GeV$, $m_{h^0}=100\, GeV$,
$m_{A^0}=200\, GeV$ and the intermediate value of
$sin\,\theta_{\mu}=0.5$. Here the lower-upper solid (dashed, small
dashed) line represents the EDM for a single-two extra dimensions,
for $\bar{\xi}^{E}_{N,\tau \mu} =10\,(30,\, 50)\, GeV$. The EDM is
slightly larger for the case that the leptons have non-zero
Gaussian profiles in the first extra dimension, compared to the
one where the leptons have non-zero Gaussian profiles in both
extra dimensions. It is observed that $d_\mu$ is weakly sensitive
to the parameter $\rho$ in the given interval, for the chosen
value of compactification scale $1/R$. In the two extra dimensions
the numerical value of $d_{\mu}$ is larger compared to the one
single extra dimension since there is a crowd of KK modes. However
the suppression exponential factor appearing in the summations
causes that the contributions do not increase extremely. The
numerical value of $d_{\mu}$ is at the order of the magnitude of
$5.0\times 10^{-22}\, (e-cm)$ for $\bar{\xi}^{E}_{N,\tau \mu}
=30\,GeV$, in the case that the leptons have non-zero Gaussian
profiles in the first extra dimension.

\begin{figure}[htb]
\vskip -3.0truein \centering \epsfxsize=6.8in
\leavevmode\epsffile{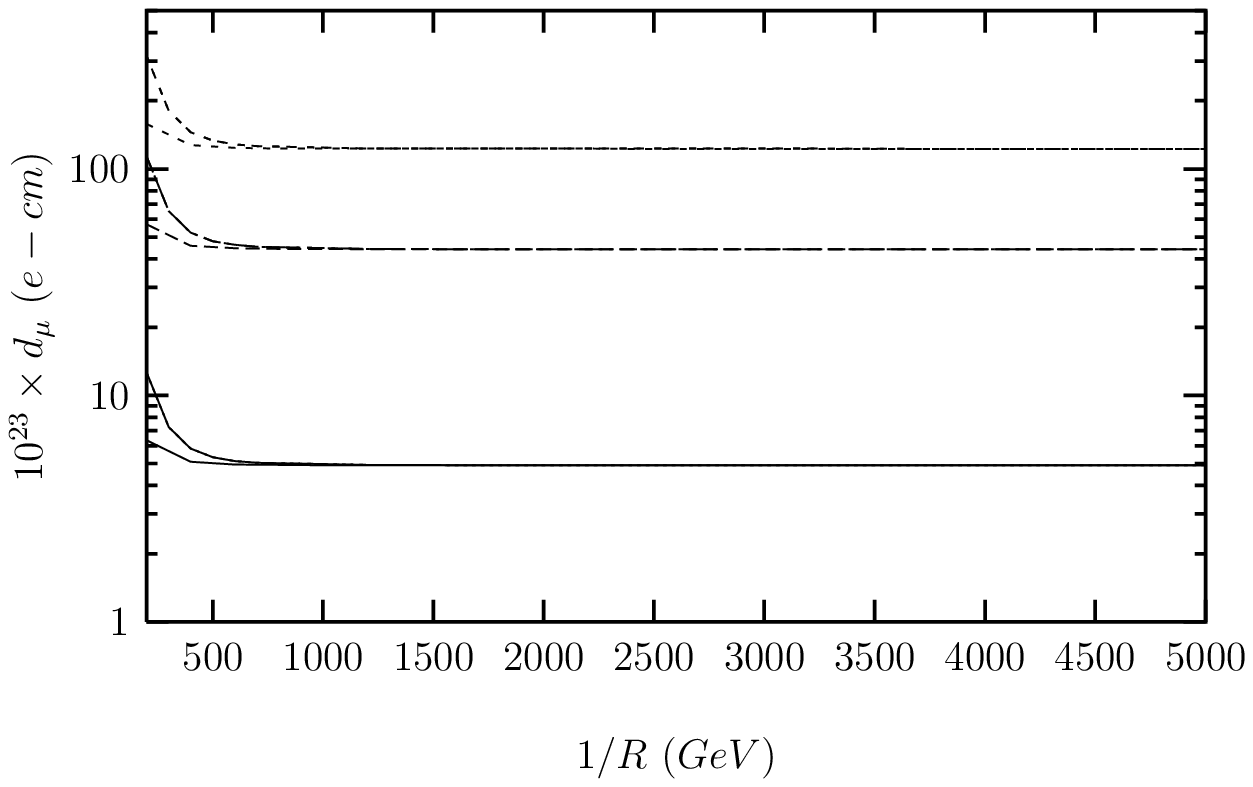} \vskip -3.0truein
\caption[]{$d_{\mu}$ with respect to the scale $1/R$, for
$\rho=0.01$, $m_{h^0}=100\, GeV$, $m_{A^0}=200\, GeV$ and the
intermediate value of $sin\,\theta_{\mu}=0.5$. Here the
lower-upper solid (dashed, small dashed) line represents the
$d_{\mu}$ for a single-two extra dimensions, for
$\bar{\xi}^{E}_{N,\tau \mu} =10\,(30,\, 50)\, GeV$.}
\label{EDMmu001Rr}
\end{figure}

Fig. \ref{EDMmu001Rr} is devoted to the EDM $d_{\mu}$ with respect
to the compactification scale $1/R$, for $\rho=0.01$,
$m_{h^0}=100\, GeV$, $m_{A^0}=200\, GeV$ and the intermediate
value of $sin\,\theta_{\mu}=0.5$. Here the lower-upper solid
(dashed, small dashed) line represents the EDM for a single-two
extra dimensions, where the leptons have non-zero Gaussian
profiles in the first extra dimension, for $\bar{\xi}^{E}_{N,\tau
\mu} =10\,(30,\, 50)\, GeV$. This figure shows that $d_\mu$ is
weakly sensitive to the compactification scale $1/R$, especially
for $1/R>500\,GeV$

\begin{figure}[htb]
\vskip -3.0truein \centering \epsfxsize=6.8in
\leavevmode\epsffile{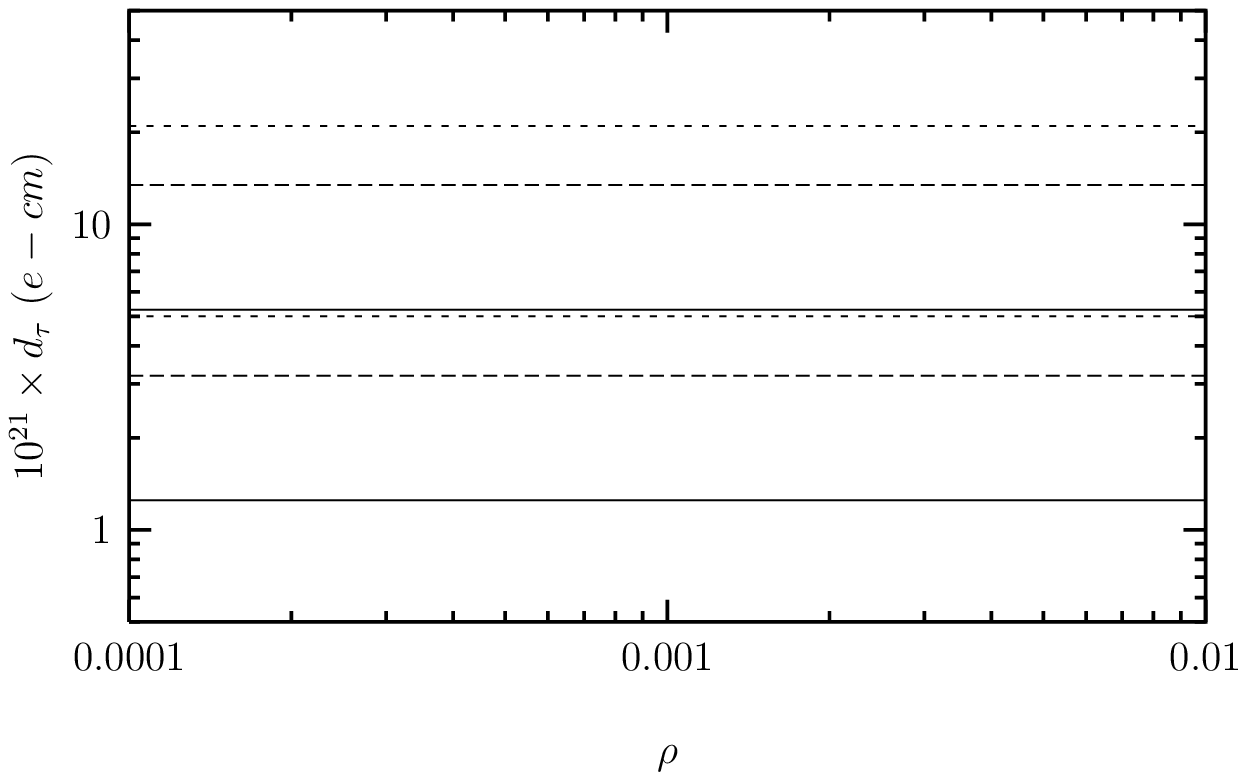} \vskip -3.0truein
\caption[]{$d_{\tau}$ with respect to the parameter $\rho$ for
$1/R=500\,GeV$, $m_{h^0}=100\, GeV$, $m_{A^0}=200\, GeV$ and the
intermediate value of $sin\,\theta_{\tau}=0.5$. Here the
lower-upper solid (dashed, small dashed) line represents the EDM
for a single-two extra dimensions, where the leptons have non-zero
Gaussian profiles in the first extra dimension, for
$\bar{\xi}^{E}_{N,\tau \tau} =50\,(80,\, 100)\, GeV$.}
\label{EDMtau500ro}
\end{figure}

In  Fig. \ref{EDMtau500ro}, we present EDM $d_{\tau}$ with respect
to the parameter $\rho$ for $1/R=500\,GeV$, $m_{h^0}=100\, GeV$,
$m_{A^0}=200\, GeV$ and the intermediate value of
$sin\,\theta_{\tau}=0.5$. Here the lower-upper solid (dashed,
small dashed) line represents the EDM for a single-two extra
dimensions, where the leptons have non-zero Gaussian profiles in
the first extra dimension, for $\bar{\xi}^{E}_{N,\tau \tau}
=50\,(80,\, 100)\, GeV$. For the case where the leptons have
non-zero Gaussian profiles in both extra dimensions, the numerical
value of $d_{\tau}$ is almost the same as the one where the
leptons have non-zero Gaussian profiles in only one extra
dimension, for two extra dimension scenario. It is shown that
$d_\tau$ is weakly sensitive to the parameter $\rho$ in the given
interval. Due to the crowd of KK modes, in the two extra
dimensions, the numerical value of $d_{\tau}$ is almost five times
larger compared to the one in the single extra dimensions. The
numerical value of $d_{\tau}$ is at the order of the magnitude of
$10^{-20}\, (e-cm)$ for $\bar{\xi}^{E}_{N,\tau \tau} =80\,GeV$, in
the two extra dimensions.

\begin{figure}[htb]
\vskip -3.0truein \centering \epsfxsize=6.8in
\leavevmode\epsffile{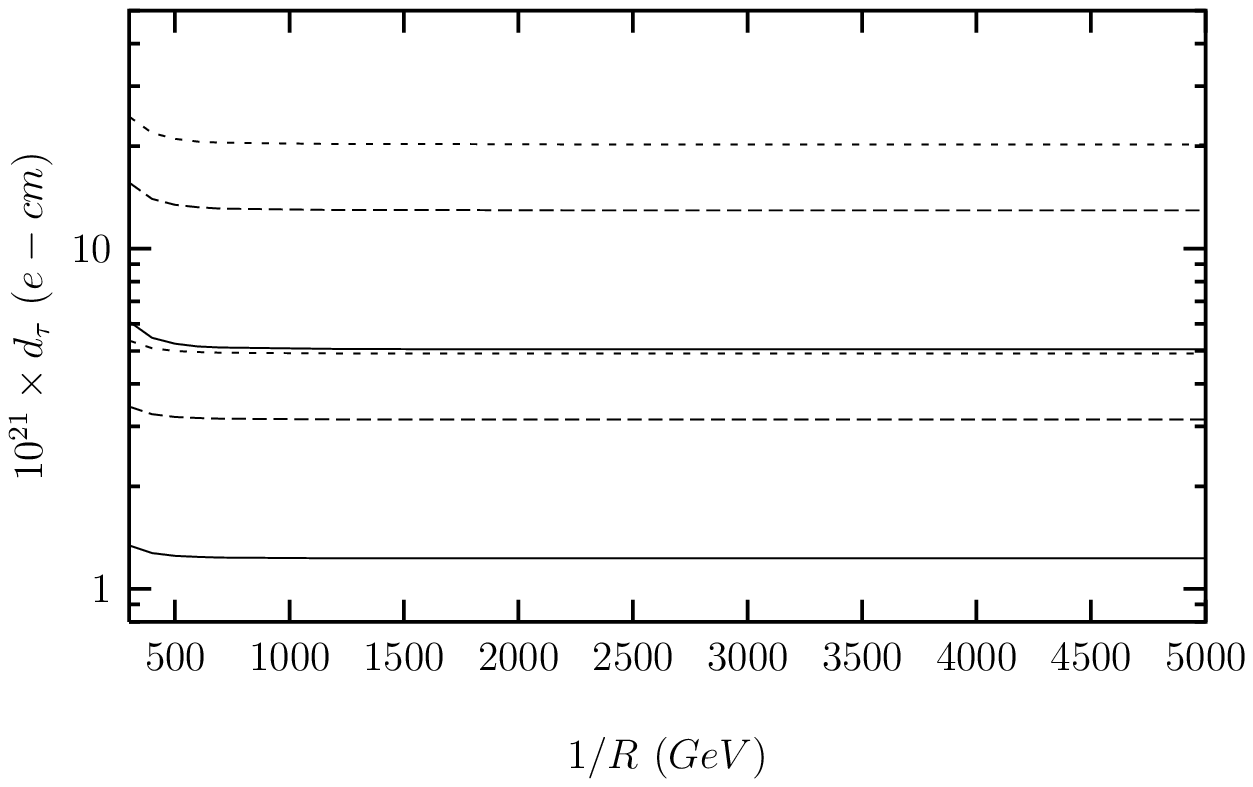} \vskip -3.0truein
\caption[]{$d_{\tau}$ with respect to the scale $1/R$, for
$\rho=0.01$, $m_{h^0}=100\, GeV$, $m_{A^0}=200\, GeV$ and the
intermediate value of $sin\,\theta_{\tau}=0.5$. Here the
lower-upper solid (dashed, small dashed) line represents the EDM
for a single-two extra dimensions, where the leptons have non-zero
Gaussian profiles in the first extra dimension, for
$\bar{\xi}^{E}_{N,\tau \tau} =50\,(80,\, 100)\, GeV$.}
\label{EDMtau001Rr}
\end{figure}

Fig. \ref{EDMtau001Rr} represents the compactification scale $1/R$
dependence of the EDM $d_{\tau}$, for $\rho=0.01$, $m_{h^0}=100\,
GeV$, $m_{A^0}=200\, GeV$ and the intermediate value of
$sin\,\theta_{\tau}=0.5$. Here the lower-upper solid (dashed,
small dashed) line represents the EDM for a single-two extra
dimensions, where the leptons have non-zero Gaussian profiles in
the first extra dimension, for $\bar{\xi}^{E}_{N,\tau \tau}
=50\,(80,\, 100)\, GeV$. For the case where the leptons have
non-zero Gaussian profiles in both extra dimensions the numerical
values of $d_{\tau}$ is almost the same as the one where the
leptons have non-zero Gaussian profiles in only one extra
dimension. Similar to the $\mu$ EDM case  $d_\tau$ is weakly
sensitive to the compactification scale $1/R$, especially for
$1/R>500\,GeV$.
\\ \\

Now we would like to summarize our results:

\begin{itemize}
\item $d_\mu$ is weakly sensitive to the parameter $\rho$ for
$\rho < 0.01$ and  the compactification scale $1/R> 500\,GeV$. Due
to the abundance of KK modes, in the two extra dimensions the
numerical value of $d_{\mu}$ is slightly larger compared to the
one in the single extra dimension. The numerical value of
$d_{\mu}$ is at most at the order of the magnitude of $5.0\times
10^{-22}\, (e-cm)$ for $\bar{\xi}^{E}_{N,\tau \mu} =30\,GeV$, in
the case that the leptons have non-zero Gaussian profiles in the
first extra dimension.
\item  $d_\tau$ is weakly sensitive to the parameter $\rho$ for
$\rho < 0.01$ and  the compactification scale $1/R> 500\,GeV$. The
crowd of KK modes in the two extra dimensions bring additional
contributions which enhance $d_{\tau}$ almost five times compared
to the one in a single extra dimension. The numerical value of
$d_{\tau}$ is at the order of the magnitude of $10^{-20}\, (e-cm)$
for $\bar{\xi}^{E}_{N,\tau \tau} =80\,GeV$, in the two extra
dimensions.
\item  The addition of the effects of the internal lepton KK modes
brings an extra dependence of the physical parameters to the scale
$1/R$, however, for the large values of the parameter $1/R$,
hopefully, these contributions do not affect the scale $1/R$
dependence of the physical parameters.
\end{itemize}

With the help of the forthcoming most accurate experimental
measurements, the valuable information can be obtained about the
existence of extra dimensions and the possibilities of Gaussian
profiles of the leptons.
\section{Acknowledgement}
This work has been supported by the Turkish Academy of Sciences in
the framework of the Young Scientist Award Program.
(EOI-TUBA-GEBIP/2001-1-8)
\newpage
\newpage
\end{document}